\documentclass[twocolumn,nofootinbib,amsmath,amssymb,aps,prd]{revtex4-1}

\usepackage{graphicx}
\usepackage[caption=false]{subfig}

\usepackage{amsmath,amssymb}
\usepackage{amsfonts,amssymb,mathrsfs}

\usepackage[bookmarks=false,pdfstartview=FitH]{hyperref}
\usepackage[all]{hypcap}
\usepackage{natbib}

\def\be{\begin{equation}}
\def\ee{\end{equation}}
\def\nn{\nonumber}
\def\f{\frac}
\def\tf{\tfrac}
\def\pl{{\rm Pl}}
\def\lp{\ell_\pl}

\def\h{\hat}

\def\dd{{\rm d}}

\def\ket{\rangle}
\def\bra{\langle}

\def\de{\delta}

\def\la{\lambda}

\def\mR{\mathcal{R}}
\def\mH{\mathcal{H}}
\def\mO{\mathcal{O}}

\def\mC{\mathcal{C}}
\def\mN{\mathcal{N}}

\def\mQ{\mathcal{Q}}
\def\su{\mathfrak{su(2)}}

\usepackage{color}

\begin{document}

\pagestyle{plain}

\title{Testing loop quantum cosmology}

\author{Edward Wilson-Ewing} \email{wilson-ewing@aei.mpg.de}
\affiliation{Max Planck Institute for Gravitational Physics (Albert Einstein Institute),\\
Am M\"uhlenberg 1, 14476 Golm, Germany, EU}

\begin{abstract}

Loop quantum cosmology predicts that quantum gravity effects resolve the big-bang singularity and replace it by a cosmic bounce.  Furthermore, loop quantum cosmology can also modify the form of primordial cosmological perturbations, for example by reducing power at large scales in inflationary models or by suppressing the tensor-to-scalar ratio in the matter bounce scenario; these two effects are potential observational tests for loop quantum cosmology.  In this article, I review these predictions and others, and also briefly discuss three open problems in loop quantum cosmology: its relation to loop quantum gravity, the trans-Planckian problem, and a possible transition from a Lorentzian to a Euclidean space-time around the bounce point.

\end{abstract}

\maketitle

\section{Introduction}
\label{s.intro}

It is notoriously difficult to test any theory of quantum gravity since any effects are typically expected to become important only near the Planck scale, which is well out of the reach of particle accelerators or even cosmic rays.  However, quantum gravity effects were likely important in the early universe at times when the space-time curvature was of the order of the inverse Planck length squared, $R \sim \lp^{-2}$, and although the cosmic microwave background (CMB) formed at a much later time, it is nonetheless possible that quantum gravity effects in the very early universe may have left a mark in primordial perturbations that could be observed in the CMB today.

Indeed, the results of high precision imaging of the CMB by the WMAP \cite{Hinshaw:2012aka} and Planck \cite{Ade:2015xua} collaborations offer the realistic hope that it may be possible to detect sub-leading quantum gravity effects if they are not too small.  In addition, there have been some surprises, with strong bounds on the tensor-to-scalar ratio \cite{Ade:2015xua} and non-Gaussianities \cite{Ade:2015ava} which were not necessarily expected from the inflationary point of view (although they by no means rule out inflation).  Perhaps these surprises are hints of something deeper that may come from quantum gravity?  In any case, observations of the early universe may well give important insights into quantum gravity.

In this review, I will focus on the predictions of loop quantum cosmology (LQC) and on the possibility of testing LQC through observations of the CMB. 

In LQC, symmetry-reduced space-times are quantized following the same procedures as loop quantum gravity (LQG).  One of the main results of LQC is that the big-bang and big-crunch singularities are resolved by quantum gravity effects \cite{Bojowald:2001xe} and are in fact replaced by a non-singular bounce \cite{Ashtekar:2006wn}.  I will review these results in Sec.~\ref{s.lqc}, with a focus on the spatially flat Friedmann-Lema\^itre-Robertson-Walker (FLRW) space-time.

More recently, there has been considerable work in determining quantum gravity corrections to the equations of motion for cosmological perturbations, with several complementary approaches having been developed, and then using these LQC-corrected equations of motion to calculate predictions that can be tested by observations of the CMB.  Quantum gravity effects in the very early universe can arise directly from the presence of a non-singular bounce (and hence the existence of a pre-bounce epoch), and also from any quantum gravity modifications to the equations of motion for the perturbations.  Of course, just like in general relativity the dynamics depend on the matter fields present and therefore so do the predictions; a number of possibilities have been studied in some detail.  In Sec.~\ref{s.perts} I will describe the three main approaches to cosmological perturbation theory developed so far in LQC, and in Sec.~\ref{s.obs} I will review the predictions of LQC in inflation, the matter bounce scenario and ekpyrosis.

I will also briefly present the cosmological constant problem from the LQC perspective in Sec.~\ref{s.lambda}, discuss some open problems in Sec.~\ref{s.open}, and end with a brief summary of the main points of this article in Sec.~\ref{s.disc}.

The conventions used in this article are the following: units are chosen so that $c=1$, while $G$ and $\hbar$ are left explicit, and the Planck length is defined as $\lp^2 =  G \hbar$.  The space-time metric is assumed to have a signature $(-,+,+,+)$, indices $a, b, c, \ldots$ refer to spatial coordinates while $i, j, k, \ldots$ are internal $\su$ indices.  The $\tau_i$ denote a basis in the $\su$ Lie algebra and satisfy $\tau_i \tau_j = \tf{1}{2} \epsilon_{ij}{}^k \tau_k - \tf{1}{4} \mathbb{I}$ with $\mathbb{I}$ being the $2 \times 2$ identity matrix.

\bigskip

\section{Homogeneous Loop Quantum Cosmology}
\label{s.lqc}

In this section I will review the theory underlying the loop quantum cosmology of homogeneous space-times in Sec.~\ref{ss.th} as well as its results and predictions in Sec.~\ref{ss.res}, focusing on the spatially flat FLRW space-time.  Readers who are only interested in the results and predictions of LQC can skip the first part of this section and go directly to Sec.~\ref{ss.res} without any loss of continuity.

\subsection{Theory}
\label{ss.th}

The key idea in LQC is to use the same fundamental variables and quantization techniques as in loop quantum gravity and apply them to cosmological space-times of interest, taking full advantage of the simplifications that arise due to the symmetries of these space-times.

Homogeneous space-times are particularly easy to handle since they have a finite number of degrees of freedom.  For example, the spatially flat FLRW space-time, with the line element
\be \label{metric}
ds^2 = - N(t)^2 dt^2 + a(t)^2 d \vec x^2,
\ee
has only one degree of freedom in its geometric sector, namely the scale factor $a(t)$.  (The lapse $N(t)$ can be freely chosen given the freedom in reparametrizing $t$.)  Assuming that the only matter content is a scalar field, then the total phase space is four-dimensional: $(a, \pi_a, \phi, \pi_\phi)$.  In the following, I will briefly review the main steps of the loop quantization of this space-time, skipping some technical details that are not necessary to understand the results.  For more details, see, e.g., \cite{Bojowald:2001xe, Ashtekar:2006wn, Ashtekar:2003hd} or the reviews \cite{Bojowald:2008zzb, Ashtekar:2011ni, Banerjee:2011qu, Agullo:2016tjh}.

The geometrical sector of the phase space can be rewritten in terms of the $\su$-valued Ashtekar-Barbero connection $A_a = A_a^i \tau_i$ and its conjugate momentum, the densitized triad $E^a_i$, as
\be \label{vars}
A_a^i = c \, (\dd x^i)_a, \qquad E^a_i = p \, \left( \f{\partial}{\partial x^i} \right)^a.
\ee
The phase space variables $c$ and $p$ are canonically conjugate%
\footnote{For the symplectic structure (and the Hamiltonian formalism in general) to be well-defined, it is necessary that integrals over the spatial slice $\Sigma$ be finite.  Due to the homogeneity of the spatial slice, there cannot be any fall-off at infinity and therefore, if the spatial slice is non-compact, it is necessary to restrict integrals to a compact region $\mathcal{V}$.  This can be seen as an infrared regulator necessary to ensure that the symplectic structure (and Hamiltonian framework) be well-defined, and it should be removed by sending $\mathcal{V} \to \Sigma$ once the equations of motion are derived.  For simplicity, here I shall choose $\mathcal{V}$ such that $\int_\mathcal{V} d^3 \vec x = 1$.  For more details on $\mathcal{V}$ see, e.g., \cite{Ashtekar:2006wn, Ashtekar:2011ni}.},
\be
\{c, p\} = \f{8 \pi G \gamma}{3},
\ee
where $\gamma$ is the Barbero-Immirzi parameter, and these variables are related to the scale factor by $p = a^2$ and%
\footnote{The relation $c = \gamma \dot{a}$ is a result of solving the equations of motion and only holds in classical general relativity.  Quantum gravity effects will modify this relation.}
$c = \gamma \dot{a}$, where the dot denotes a derivative with respect to proper time (i.e., the time coordinate for $N=1$).

The holonomy of $A_a^i$ along a line segment parallel to $x^j$ depends on the length $\lambda$ of the line segment as
\begin{align}
h_j(\lambda) &=
\exp \left[ \int_0^\lambda \!\! d x^j A_a \left( \f{\partial}{\partial x^j} \right)^a \right] \nn \\
&= \cos \f{\la c}{2} \mathbb{I} + 2 \sin \f{\la c}{2} \tau_j,
\end{align}
where there is no sum over $j$ on the first line.  (It is sufficient to only consider holonomies along edges parallel to the $x^j$ due to the homogeneity of the spatial slice \cite{Ashtekar:2003hd}.)  Note that the length $\lambda$ is calculated with respect to the so-called fiducial (spatial) metric whose line element is $d\mathring{s}^2 = d\vec x^2$, as can be seen from the measure $dx^j$ inside the integral on the first line.  An important point here is that the dependence on $c$ of these `straight' holonomies can be expressed entirely in terms of complex exponentials of $c$.

Then, following loop quantum gravity, the elementary operators in LQC are the surface area operator corresponding to $p$ and operators corresponding to complex exponentials of $c$, which are sufficient to define operators corresponding to the SU(2)-valued holonomies along straight line segments.  It is convenient to use the basis $|p\ket$ for the gravitational kinematical Hilbert space $H_g$, in which case these operators act as:
\be
\hat p |p\ket = p |p\ket,
\ee
\be \label{mN}
\mN(\lambda) |p\ket := \widehat{e^{i \lambda c}} |p\ket = |p - 8 \pi \gamma \lp^2 \lambda / 3 \ket.
\ee
The inner product between two such basis vectors is
\be
\bra p | \tilde p \ket = \delta_{p, \tilde p},
\ee
where $\delta_{p, \tilde p}$ is the Kronecker delta, not the Dirac delta distribution.  States $\psi(p) = \sum_p \psi_p |p\ket$ in $H_g$ are those that are normalizable with this inner product.  An important consequence of this inner product is that it is not possible to take the limit $\lambda \to 0$ in $-i[ \mN(\lambda) - 1] / \lambda$ to obtain an operator $\hat c$ since this limit is not well-defined on the kinematical Hilbert space: there is no connection operator, only a holonomy operator for finite $\lambda$.  As a technical remark, note that since the limit of $\lambda \to 0$ of $\mN(\lambda)$ is not well-defined, $\mN(\lambda)$ is not weakly continuous in $\lambda$ and therefore the Stone-von Neumann uniqueness theorem is not applicable: this is one reason that LQC does not give the same physical predictions as Wheeler-de Witt quantum cosmology models.  (Another reason is that the Hamiltonian constraint operator is expressed in terms of holonomies of $c$, as explained below.)

Then, the total kinematical Hilbert space is $H_k = H_g \otimes H_m$, where the matter kinematical Hilbert space consists of square-integrable functions $\psi(\phi)$ with respect to the Lebesgue measure on $\phi$, and the elementary operators are
\be
\hat \phi \, \psi(\phi) = \phi \, \psi(\phi), \qquad
\widehat{\pi_\phi} \, \psi(\phi) = -i \hbar \f{d \psi(\phi)}{d\phi}.
\ee

The classical dynamics are generated by the Hamiltonian constraint
\be \label{cl-ham}
\mC_H = \int \left[ N \mH + N^a \mH_a + \Lambda^i \mathcal{G}_i \right] \approx 0,
\ee
and due to the gauge-fixing chosen in \eqref{metric} and \eqref{vars}, the diffeomorphism constraint $\mH_a$ and the Gauss constraint $\mathcal{G}_i$ are already automatically satisfied.  The `$\approx 0$' in \eqref{cl-ham} denotes that $\mC_H$ is a constraint and must vanish for physical solutions.  Furthermore, since the integral over a homogeneous spatial manifold is trivial, this gives $\mC_H = N \mH$ where the scalar constraint $\mH$, for a spatially flat FLRW space-time with a massless scalar field, is simply
\be
\mH = - \f{E^a_i E^b_j}{8 \pi \gamma^2 G \sqrt{q}} \, \epsilon^{ij}{}_k F_{ab}{}^k + \f{\pi_\phi^2}{2 \sqrt{q}},
\ee
where $F_{ab}{}^k = 2 \partial_{[a} A_{b]}^k + \epsilon_{ij}{}^k A_a^i A_b^j$ is the field strength of the connection $A_a^i$ and $q = |p|^3$ is the determinant of the spatial metric.

To define an operator corresponding to the Hamiltonian constraint, it is necessary to first define an operator corresponding to $F_{ab}{}^k$.  (Operators corresponding to $E^a_i \sim p$ and $\pi_\phi$ are already defined%
\footnote{It is also necessary to define inverse triad operators since the state $|p=0\ket \in H_g$ is an eigenstate of $\h p$ with eigenvalue zero.  There is considerable ambiguity in the choice of inverse triad operators, but in non-compact spaces all known inverse triad operators in LQC tend to the trivial inverse triad operator
\be
\widehat{\f{1}{p}} \, |p\ket =
\begin{cases}
0 ~~ &{\rm if}~p = 0, \\
\f{1}{p} \, |p\ket ~~&{\rm otherwise.}
\end{cases} \nn
\ee
in the limit that the fiducial cell is removed \cite{Singh:2013ava}.  Therefore, all known inverse triad operators in LQC can only have a non-trivial effect in compact spaces, and even in that case their effect is small so long as the spatial volume of the space-time at the bounce point is large compared to $\lp^3$ (which it typically is).}.)
This is non-trivial since there only exist operators corresponding to holonomies of $A_a^i$, but not to $A_a^i$ itself.  However, there is a simple and natural solution: use Wilson loops to define the field strength, i.e., by taking the holonomy of $A_a^i$ around a small closed loop.  In standard quantum field theory, one would take the limit of the area of the loop going to zero.

However, this limit cannot be taken in LQC since $\lim_{\lambda\to0} \mN(\lambda)$ does not exist.  In fact, it is not natural to take this limit in LQC, for the reason that LQC is based on loop quantum gravity where the spectrum of the area operator is discrete with a minimum non-zero eigenvalue $\Delta \lp^2$ (with $\Delta = 4 \sqrt 3 \pi \gamma$), and therefore the limit ${\rm Area} \to 0$ does not exist in LQG either.

Instead, it is more natural to construct the field strength operator by taking the holonomy of $A_a^i$ around the minimal loop possible according to loop quantum gravity, i.e., by setting the physical area of the loop to $\Delta \lp^2$:
\begin{align} \label{field}
\hat F_{ab}{}^k &= 2 \, {\rm Tr} \left[ \f{h_j^{-1}(\bar\mu)h_i^{-1}(\bar\mu)h_j(\bar\mu)h_i(\bar\mu)}{\bar\mu^2} \, \tau^k \right] (dx^i)_a (dx^j)_b \nn \\
&= \f{\sin^2 \left(\sqrt{\Delta \lp^2 / |p|} c \right)}{\Delta \lp^2 / |p|} \epsilon_{ij}{}^k (dx^i)_a (dx^j)_b.
\end{align}
Note that the length $\lambda$ in \eqref{mN} is measured with respect to the fiducial metric, and the area $\Delta \lp^2$ encircled by the holonomy is a physical area evaluated with respect to the metric \eqref{metric}.  For this reason, it is necessary to set the length of each holonomy to be $\bar\mu = \sqrt{\Delta \lp^2 / |p|}$ to ensure that the physical length of each side of the square is $\sqrt{\Delta} \lp$, as required.

The last step is to define the action of complex exponentials of $\bar\mu c$ on the kinematical Hilbert space; since $\bar\mu$ depends on $p$, the action is not the same as the operator \eqref{mN}.  The action of this new operator follows from the fact that the quantity canonically conjugate to $b = \bar\mu c$ is $V = {\rm sgn}(p) |p|^{3/2}$ (with $\{b, V\} = 4 \pi \gamma \sqrt{\Delta} G\lp$), and so
\be
\widehat{e^{\pm ib}} |V\ket = |V \mp 4 \pi \gamma \sqrt\Delta G\lp\ket,
\ee
where $|V\ket$ denotes the basis vectors in the volume representation (which is just a relabeling of the area representation basis vectors $|p\ket$).

Then, given the definition of the field strength operator \eqref{field} the Hamiltonian constraint operator, for a given choice of the lapse $N$, follows.  The exact action of the Hamiltonian constraint depends on a number of choices, including factor-ordering choices, the choice of the lapse, and the definition of inverse triad operators.  However, no matter what choices are made, the action of the Hamiltonian constraint operator on states in the physical Hilbert space (i.e., the states annihilated by the Hamiltonian constraint operator, $\hat \mC_H \Psi = 0$) has the following form:
\begin{align} \label{qham}
- \hbar^2 \partial_\phi^2 \Psi(V, \phi) = \:& C_{V+} \Psi(V_+, \phi) + C_0 \Psi(V, \phi) \nn \\
& \quad + C_{V-} \Psi(V_-, \phi),
\end{align}
where $V_\pm = V \pm  8 \pi \gamma \sqrt\Delta G\lp$ and the exact form of the $C_V$ terms depends on the quantization ambiguities.  This operator is essentially self-adjoint \cite{Kaminski:2007gm}.  The explicit form of the $C_V$ is not particularly illuminating; for specific expressions of the $C_V$ given some specific choices for the factor-ordering, the lapse, and the inverse triad operators, see, e.g., \cite{Ashtekar:2006wn, Ashtekar:2007em, MartinBenito:2009aj}.

Rather, the main points are that the quantum equation of motion \eqref{qham}: (i) is a difference equation in $V$ rather than a differential equation, this is a consequence of the discrete nature of quantum geometry in LQG, and (ii) gives the evolution of the quantum cosmology wave function in a relational sense, where the scalar field $\phi$ acts as a relational clock.  Thus, the main (Dirac) observable of interest is the volume $V$ evaluated at an instant of relational time $\phi_o$.

\subsection{Results and Predictions}
\label{ss.res}

To recap, in the LQC of the spatially flat FLRW space-time with a massless scalar field $\phi$, the wave function is usually studied in the representation $\Psi(V, \phi)$, where $V = a^3$, and in addition the scalar field $\phi$ can act as a relational clock with respect to which the wave function is evolved.

The quantum dynamics of LQC can be studied by choosing an initial state $\Psi(V, \phi_o)$ at some instant $\phi_o$ of relational time and numerically evolving it using the LQC Hamiltonian constraint operator \eqref{qham}.  This was first done for initial states sharply peaked around a classical solution to the Friedmann equations at a sufficiently small energy density so that quantum gravity effects are initially negligible.  The results of numerically solving \eqref{qham} for such initial conditions are the following \cite{Ashtekar:2006wn}: (i) the wave function remains sharply peaked throughout the entire evolution, (ii) the wave packet follows the classical Friedmann trajectory very closely so long as the matter energy density remains small compared to the Planck scale, and (iii) when the matter energy density nears the Planck scale, the wave packet departs from the classical theory and bounces at a large but finite critical energy density of the matter field $\rho_c \sim \rho_{\rm Pl}$.  Once the energy density decreases sufficiently far below the Planck scale, then the wave packet recommences to follow a classical Friedmann trajectory once more.

While numerical studies first considered states that are sharply peaked around classical solutions, a number of more recent studies have shown that a large class of widely spread states that do not have a nice semi-classical limit also bounce, with the same upper bound on the expectation value of the matter energy density (in fact, states with a large spread typically bounce at a lower expectation value of the energy density than sharply peaked states) \cite{Diener:2013uka, Diener:2014mia, Diener:2014hba}.  Furthermore, for a certain lapse and with certain factor-ordering choices, it is possible to obtain a Hamiltonian constraint operator which is exactly soluble, and in that case it can be shown analytically that the bounce is generic and that the energy density of the scalar field is bounded above by $\rho_c$ \cite{Ashtekar:2007em}.  Finally, the quantization ambiguities in the definition of the Hamiltonian constraint operator have also been studied numerically, with the result that the qualitative results, including the bounce and the upper bound on the energy density, hold for all possibilities considered in the literature \cite{MenaMarugan:2011me}.

Thus, one of the most important results in LQC, applied to a spatially flat FLRW space-time with a massless scalar field, is that the big-bang and big-crunch singularities of the spatially flat FLRW space-time are generically resolved and are replaced by a bounce.

As mentioned above, if the state is sharply peaked at an initial moment of relational time $\phi_o$ then it will remain sharply peaked throughout its evolution.  In other words, if quantum fluctuations are initially small then they remain small---this is due to the global observables of interest in quantum cosmology like the total spatial volume being heavy degrees of freedom where quantum fluctuations do not grow significantly \cite{Rovelli:2013zaa}.  In this case, since $\bra \mO^2 \ket \approx \bra \mO \ket^2$ for any observable $\mO$, it is sufficient to study the dynamics of the expectation values of the observables of interest.  This gives the effective Friedmann equations of LQC \cite{Ashtekar:2006wn, Taveras:2008ke}
\be \label{fr1}
H^2 = \f{8 \pi G}{3} \rho \left(1 - \f{\rho}{\rho_c}\right),
\ee
\be \label{fr2}
\dot{H} = -4 \pi G (\rho + P) \left(1 - \f{2 \rho}{\rho_c}\right),
\ee
\be \label{cont}
\dot\rho + 3 H (\rho + P) = 0,
\ee
where $H = \dot{a}/a$ is the Hubble rate, $\rho$ and $P$ denote the energy density and the pressure of the matter field respectively, $\rho_c \sim \rho_{\rm Pl}$ is the critical energy density of LQC, and dots denote derivatives with respect to proper time.

From these equations, it is clear that there is a bounce when $\rho = \rho_c$, and also that quantum gravity effects are negligible when $\rho \ll \rho_c$.  Furthermore, while the equations of motion for the gravitational degrees of freedom are modified by quantum gravity corrections, the continuity equation for the matter field remains unchanged: LQC effects only arise in the geometrical sector of the theory.

Finally, from these effective equations another important point is made clear: quantum gravity effects, for sharply-peaked states in LQC, become important when the energy density (equivalently, when the space-time curvature) approaches the Planck scale.  Note that the spatial volume of the space-time may be (and typically will be) very large compared to the Planck scale when the bounce occurs.  (In fact, it will be infinite for non-compact spaces.)  Therefore, the relevant length scale that determines the amplitude of LQC effects is the radius of the space-time curvature, not the radius of the spatial volume.  (In the case that the spatial volume nears $\lp^3$, then quantum fluctuations will become important and generate additional quantum gravity effects.  However, this should not be confused with the LQC effects that cause the bounce.)

So far, I have reviewed the results of LQC as applied to the spatially flat FLRW space-time with a massless scalar field.  Many other homogeneous space-times have also been studied in LQC, including: the closed and open FLRW space-times \cite{Ashtekar:2006es, Szulc:2006ep, Vandersloot:2006ws} the Bianchi type I, type II and type IX space-times \cite{Ashtekar:2009vc, Ashtekar:2009um, WilsonEwing:2010rh}, and the Kantowski-Sachs space-time \cite{Bohmer:2007wi, Corichi:2015xia} (which corresponds to the interiour of a Schwarzschild black hole).  Different matter fields have also been studied, namely Maxwell fields \cite{Pawlowski:2014fba} and inflationary fields \cite{aps-infl}, and it has also been shown how to include either a positive or a negative cosmological constant \cite{Kaminski:2009pc, Pawlowski:2011zf, Bentivegna:2008bg}.

In all of these cases, the big-bang and big-crunch singularities are resolved by quantum gravity effects.  Furthermore, for FLRW space-times a numerical analysis of the quantum dynamics shows that sharply peaked states bounce at $\rho_c$ and there again exist effective Friedmann equations that provide an excellent approximation to the full quantum dynamics of sharply peaked states \cite{Ashtekar:2006es, Szulc:2006ep, Vandersloot:2006ws, Pawlowski:2014fba, aps-infl, Kaminski:2009pc, Pawlowski:2011zf, Bentivegna:2008bg}.

For the Bianchi and Kantowski-Sachs space-times, the Hamiltonian constraint operator is significantly more complicated and has not yet been solved numerically.  Nonetheless, there exist effective equations for the Bianchi and Kantowski-Sachs space-times as well, and analytic and numerical studies of the effective equations find that a Planck-scale bounce replaces the classical big-bang singularity in these space-times also \cite{Corichi:2009pp, Gupt:2011jh, Gupt:2012vi, Corichi:2012hy, Singh:2013ava, Corichi:2015ala, Chiou:2008eg}.  (Note that the observables of interest in Bianchi space-times are global observables, and therefore they are heavy degrees of freedom so long as all three $p_i \sim a_j a_k$ remain much larger than $\lp^2$.  As a result, quantum fluctuations will not play an important role in states where the fluctuations are initially small and $p_i \gg \lp^2$ at all times, and then for these states the effective equations will provide a good approximation to the full quantum dynamics \cite{Rovelli:2013zaa}.)

Furthermore, the effective equations also show that for sharply-peaked states of FLRW, Bianchi type I and Kantowski-Sachs space-times all strong singularities are resolved by quantum gravity effects \cite{Singh:2009mz, Singh:2010qa, Singh:2011gp, Saini:2016vgo}.

Finally, while there are a number of quantization ambiguities in LQC, the main results---namely, the generic bounce and the reliability of the effective equations for initially sharply peaked states---are robust and are not affected by factor-ordering choices \cite{MenaMarugan:2011me}, changes in the definition of the field strength operator \cite{Corichi:2011pg, Singh:2013ava}, and even changes in the elementary variables: similar results are obtained if one uses self-dual variables rather than the Ashtekar-Barbero variables \cite{Achour:2014rja, Wilson-Ewing:2015lia, Wilson-Ewing:2015xaq}.

\section{Three Approaches to Cosmological Perturbation Theory}
\label{s.perts}

There are three main approaches to cosmological perturbation theory that have been developed in LQC: effective constraints, hybrid quantization, and separate universes.  In each case, the goal is to determine LQC effects on linear cosmological perturbations on a spatially flat FLRW background space-time, typically with a scalar field $\phi$ as the matter content.

Specifically, in general relativity it is the Mukhanov-Sasaki equation that determines the dynamics of scalar perturbations \cite{Mukhanov:1990me},
\be \label{class-ms}
v_k'' + k^2 v_k - \f{z_s''}{z_s} v_k = 0,
\ee
where the $k^{th}$ Fourier mode of the Mukhanov-Sasaki variable is related to the co-moving curvature perturbation $\mR$ by $v_k = z_s \mR_k$ and $z_s = a \dot{\phi} / H$ is a function that depends on the background evolution.  Primes denote derivatives with respect to conformal time $\eta$, i.e., the time coordinate when $N=a$.  (For completeness, the equation of motion for the tensor perturbations $h_k$ is obtained by replacing $v_k$ by $\mu_k = a h_k$ and $z_s$ by $a$ in \eqref{class-ms}.)  The main aim of each of the three approaches to cosmological perturbation theory in LQC is to determine what modifications due to LQC, if any, should appear in these equations of motion for cosmological perturbations, and then use these results to calculate observational consequences of LQC effects.

The three frameworks handle cosmological perturbations in different ways.  The effective constraint approach is based on effective equations, but without constructing or knowing the underlying quantum theory.  On the other hand, the hybrid quantization approach is based on a well-defined quantum theory, with a loop quantization for the background variables and a Fock quantization for the perturbative degrees of freedom.  Finally, the separate universe approach gives a loop quantization of both the background and long-wavelength scalar perturbations, but ignores short-wavelength perturbations.  It remains to extend these results to higher order in perturbation theory to calculate LQC effects on non-Gaussianities, and also to study perturbations on a spatially curved and/or anisotropic background space-time.

In this section I will briefly review these three approaches, focusing on their conceptual underpinnings and main results, and pointing out the assumptions underlying each.  For a more detailed introduction of these approaches to cosmological perturbation theory in LQC, see the reviews \cite{Barrau:2013ula, Ashtekar:2015dja, Wilson-Ewing:2015sfx}.  For predictions derived from these frameworks, see Sec.~\ref{s.obs}.

It is important to state that none of these approaches is as robust as LQC for homogeneous space-times.  It is still not known how to fully extend the results reviewed in Sec.~\ref{ss.th} to allow for inhomogeneities.  Indeed, the three frameworks developed so far all avoid (in different ways) the difficult problem of performing a loop quantization of all degrees of freedom in an inhomogeneous space-time.

\subsection{Effective Constraints}
\label{ss.eff}

From a phenomenological perspective, the effective constraint approach is in large part motivated by the high accuracy of the LQC effective Friedmann equations describing the full quantum dynamics of homogeneous space-times, even at the bounce point, for states with small quantum fluctuations.  The hope here is that similar effective equations will exist and be equally accurate for cosmological perturbations as well, for example an LQC effective Mukhanov-Sasaki equation.

The challenge is to find the correct effective equations without knowing the underlying quantum theory.  The procedure followed in this approach is to take the classical scalar and diffeomorphism constraints of general relativity in Ashtekar-Barbero variables (typically the Gauss constraint is gauge-fixed, as in homogeneous LQC) for the spatially flat FLRW space-time with linear perturbations, and then allow for a large class of possible modifications motivated by LQC---typically holonomy or inverse triad effects.

Schematically, for the case of holonomy corrections, each time the connection variable $c$ or its perturbation $\de c$ appears in one of the constraints, it is replaced by some function $f_i(c)$ or $g_i(\de c)$ which is meant to encode the effects due to holonomy corrections in LQC (with these functions potentially different for each appearance of $c$ or $\de c$ in the classical constraints).  Similarly, for inverse triad corrections, each time an inverse power of $p = a^2$ appears in the classical constraints, it is replaced by some function $h_i(p)$, which is meant to encode inverse triad effects from the quantum theory. Of course, in order to recover general relativity in the classical limit, it is necessary to require that $f_i(c) \to c$ and $g_i(\de c) \to \de c$ when the curvature is small and $h_i(p) \to p^{-1}$ when $p$ is large.

Nonetheless, there is clearly a great deal of freedom in the choices of the `correction' functions.  However, an important condition---necessary to obtain a consistent theory---is that the constraints have an anomaly-free Poisson algebra.  It turns out that this requirement strongly restricts the form that these correction functions can take.  (Note that the form of the constraint algebra may change.  What is important is that the constraint algebra closes, not that it have a specific form.  In fact, in LQC the constraint algebra will typically be modified by holonomy or inverse triad corrections.)

This was first done for inverse triad corrections in the limit that the corrections be small and that there exist a perturbative expansion for them \cite{Bojowald:2006tm, Bojowald:2008gz, Bojowald:2008jv}.  In this case, the anomaly-free condition strongly restricts the possible forms of LQC inverse triad effects on the dynamics of scalar and tensor perturbations in the effective constraint formalism.

Holonomy corrections were considered next.  Holonomy corrections are particularly important in LQC, since the occurrence of the bounce in homogeneous LQC is entirely due to holonomy corrections.  Interestingly, given some reasonable assumptions on the form of the correction functions, the equations of motion for cosmological perturbations with holonomy corrections are uniquely determined by the requirement that the constraint algebra be free of quantum anomalies \cite{Cailleteau:2011kr, Cailleteau:2012fy}.  Furthermore, for holonomy corrections the modifications to the equations of motion for cosmological perturbations are particularly simple.  For scalar perturbations, the LQC effective Mukhanov-Sasaki equation with holonomy corrections is \cite{Cailleteau:2011kr}
\be \label{ef-sc}
v_k'' + \left(1 - \f{2 \rho}{\rho_c}\right) k^2 v_k - \f{z_s''}{z_s} v_k = 0,
\ee
and for tensor perturbations, holonomy corrections are captured by the effective equation \cite{Cailleteau:2012fy}
\be \label{ef-te}
\tilde \mu_k'' + \left(1 - \f{2 \rho}{\rho_c}\right) k^2 \tilde \mu_k - \f{z_t''}{z_t} \tilde \mu_k = 0,
\ee
where $\tilde \mu_k$ is related to the tensor perturbation $h_k$ by $\tilde \mu_k = z_t h_k$ and $z_t = a / \sqrt{1 - 2 \rho / \rho_c}$.

These equations can be used to quantitatively study quantum gravity effects in the early universe on cosmological perturbations and hence on structure formation.

Building on these results, it has also been shown that holonomy and inverse triad effects can be included simulataneously in the effective framework \cite{Cailleteau:2013kqa}, and a discussion on how gauge transformations are affected by modifications to the classical constraints can be found in \cite{Cailleteau:2011mi}.

In addition, the specific form of the constraint algebra when including holonomy corrections has lead to some interesting speculation concerning a possible signature change around the bounce point in LQC.  At this time, more work is needed to determine whether this interpretation of the constraint algebra is correct or not.  I will return to this question in more detail later in Sec.~\ref{ss.sig} as it is, in my opinion, one of the important open problems in LQC.

However, the effective equations \eqref{ef-sc} and \eqref{ef-te} have an important drawback in that they ignore quantum fluctuations.  (The effective framework can be extended to allow for quantum fluctuations by including higher order moments in the observables \cite{Bojowald:2012xy}, but this extension has not yet been done for perturbations.)  This is not a problem if one is interested in the dynamics of heavy degrees of freedom, but quantum fluctuations cannot be ignored when considering light degrees of freedom.  A simple calculation shows that quantum fluctuations are expected to become important and cannot be ignored when the physical wavelength of the perturbation modes of interest is of the order of $\lp$ \cite{Rovelli:2013zaa}.  Therefore, this suggests that the effective constraint approach reviewed here is a good approximation for long-wavelength modes (as compared to $\lp$), but will likely fail when applied to modes with a wavelength shorter than $\lp$.

In other words, there is a trans-Planckian problem in the effective constraint approach to perturbations in LQC: this approach cannot be used to study perturbations whose wavelength is shorter than the Planck length.  For this reason, it is not surprising that when these equations of motion are used to evolve trans-Planckian modes through the bounce, the result is a power spectrum that is ruled out by observations (even if the bounce is followed by an inflationary epoch) \cite{Bolliet:2015raa}.  This is simply the consequence of using the equations of motion \eqref{ef-sc} and \eqref{ef-te} outside of their regime of validity.

Nonetheless, despite the effective constraint approach breaking down whenever quantum fluctuations become important (and in particular for trans-Planckian modes), the equations motion are expected to hold in many settings of cosmological interest and may give important insights into quantum gravity effects in the early universe.

\subsection{Hybrid Quantization}
\label{ss.hyb}

The second framework to be developed for cosmological perturbation theory in LQC goes by the names of hybrid quantization \cite{FernandezMendez:2012vi, Fernandez-Mendez:2013jqa, Gomar:2014faa, Gomar:2016rso} and the dressed metric framework \cite{Agullo:2012fc, Agullo:2013ai, Agullo:2016hap}.  While there are some differences between these two approaches, the basic idea is the same and the differences are negligible at a phenomenological level when considering small perturbations \cite{deBlas:2016puz}.  For these reasons, these two approaches will be considered together here.

The idea underpinning the hybrid quantization is to treat the background and perturbative degrees of freedom differently, namely by performing a loop quantization of the FLRW background and a Fock quantization of the perturbative degrees of freedom.  This builds on earlier studies of both the Gowdy space-time in LQC \cite{MartinBenito:2008ej, Garay:2010sk, MartinBenito:2010bh, MartinBenito:2010up} (which can be viewed as gravitational wave inhomogeneities in one spatial dimension on a Bianchi I background) and also of a test (inhomogeneous) scalar field on an LQC background \cite{Ashtekar:2009mb}, where this type of hybrid quantization was first developed.

Thus, the basic assumption of the hybrid quantization approach to cosmological perturbation theory is that a Fock quantization is appropriate for the perturbations.  While quantum gravity effects are known to be important for the background FLRW space-time in LQC, it is assumed that these quantum gravity effects do not directly modify the quantum equations of motion for the perturbations; instead, quantum gravity effects present in the background space-time percolate to the perturbations through their equations of motion which depend on the dynamics of the background space-time.  Note that quantum fluctuations, which are expected to be important for trans-Planckian modes, are fully included in the Fock quantization.

While this approximation may initially appear quite drastic, lessons from homogeneous LQC suggest that it is reasonable.  Specifically, for sharply-peaked states in homogeneous LQC, quantum gravity effects only become important when the energy density of the matter field (or in the anisotropies) becomes comparable to the Planck scale (so long as the spatial volume remains much larger than $\lp^3$, which it typically is, even at the bounce).  Therefore, if the energy density in the perturbations always remains small compared to the Planck scale (and it does if the perturbations remain linear, this has been checked explicitly in inflationary models \cite{Agullo:2013ai}) then quantum gravity effects acting directly on the perturbations may indeed be negligible.

Also, in the hybrid quantization it is assumed that perturbations can have an arbitrarily short wavelength: there is no wavelength cut-off for the cosmological perturbations.  At this time, it is not yet clear if this assumption is justified.  A minimal wavelength for perturbations might be expected if there is a minimal non-zero eigenvalue of the length operator in loop quantum gravity, while if there is no minimal non-zero eigenvalue (i.e., if there exist eigenvalues arbitrarily close to 0) then no length cutoff should exist.  However, since the (discrete) spectrum of the length operator is unknown due to its complexity \cite{Thiemann:1996at, Bianchi:2008es, Ma:2010fy}, this question remains unanswered for now.  Thus, in the hybrid quantization approach to cosmological perturbation theory in LQC, the trans-Planckian problem is not directly addressed but rather is avoided by assuming that perturbations can have an arbitrarily small wavelength.

The hybrid quantization gives a fully quantum treatment of cosmological perturbations in LQC, a quantum field theory on a quantum background, unlike the effective constraint framework.  Furthermore, an important result is that the quantum dynamics of the perturbations only depend on a small number of the quantum properties of the background.  In fact, the dependence is so simple that the full quantum dynamics of the perturbations can be rewritten as a quantum field theory on a `dressed' background space-time, where the dressing contains the information of the few quantum properties of the background that affect the dynamics of the perturbations \cite{Agullo:2012fc, Agullo:2013ai}.  To be specific, the quantum equations of motion for scalar perturbations in the dressed metric framework are \cite{Agullo:2013ai}
\be \label{sc-hybrid}
\hat\mQ_k'' + 2 \f{\tilde a'}{\tilde a} \hat\mQ_k' + \left( k^2 + \f{\tilde a''}{\tilde a} - \f{\tilde u''}{\tilde u} \right) \hat \mQ_k = 0,
\ee
where $\mQ = v / a = z_s \mR / a$ and $u = a \sqrt{3 (1 + w_{eff}) / 8 \pi G}$, with $w_{eff} = [\dot\phi^2 / 2 - V(\phi)] / [\dot\phi^2 / 2 + V(\phi)]$ the (time-dependent) effective equation of state of the scalar field, and
\be
\tilde a^4 = \f{\bra \hat H_o^{-1/2} \, \hat a^4 \, \hat H_o^{-1/2} \ket}{\bra H_o^{-1} \ket},
\ee
\be
\tilde u = \f{\bra \hat H_o^{-1/2} \, \hat a^2 \, \hat u \, \hat a^2 \, \hat H_o^{-1/2} \ket}
{\bra \hat H_o^{-1/2} \, \hat a^4 \, \hat H_o^{-1/2} \ket},
\ee
encode the expectation values of the scale factor and $u$ as well as some information about quantum fluctuations in the scale factor.  Here $\hat H_o$ is the positive frequency LQC Hamiltonian for the background degrees of freedom with respect to the relational time variable $\phi$ or, in other words, $H_o$ is the negative square root of the operator acting on $\Psi(V, \phi)$ in the right-hand side of \eqref{qham}.

A technical point here is that, as explained above, the equation of motion \eqref{sc-hybrid} follows from a Fock quantization of the Hamiltonian for the perturbations, and this Hamiltonian of course comes from classical general relativity.  However, the function $z_s$ appearing in the Hamiltonian as $z_s''/z_s$ is not uniquely determined classically, since the Friedmann equations (of general relativity) can be freely used to rewrite $z_s$ in different equivalent forms.  Since initial conditions are typically imposed at the bounce point in the hybrid quantization approach (as shall be discussed shortly), in this case the usual function $z_s = a \dot{\phi} / H$ is problematic since it diverges at the bounce where $H=0$.  Classically, the function $u = a \sqrt{3 (1 + w_{eff}) / 8 \pi G}$ is equivalent to $z_s$ by the scalar constraint of the background space-time (which is equivalent to the classical Friedmann equation), and is better suited to the hybrid quantization framework since it doesn't diverge at the bounce point.  For this reason $u$ is typically used in the dressed metric approach, as can be seen in \eqref{sc-hybrid}.  However, $u$ and $z_s$ are not equivalent in LQC since the effective Friedmann equation \eqref{fr1} is modified by LQC effects, and it is not known whether using $u$ instead of $z_s$ could lead to substantially different predictions or not.

The equations of motion for the tensor modes $\hat h_k$ are obtained simply by replacing $\hat\mQ$ by $\hat h$ and $\tilde u$ by $\tilde a$ in \eqref{sc-hybrid} \cite{Agullo:2012fc}.  Importantly, due to the form of these quantum equations of motion, the well known standard techniques of quantum field theory on a curved space-time can be used; in particular, it is possible to define $n^{th}$-order adiabatic states and renormalize observables of interest like the energy density of perturbations.

Finally, in this framework, quantum vacuum initial conditions are typically imposed at (or near) the bounce point, based on the following heuristic argument: the bounce is caused by gravity becoming repulsive at very high energy densities, and if this repulsive `force' acts also on the perturbations it might be expected that perturbations would be smoothed out to be as `small' as possible, hence justifying setting quantum vacuum initial conditions at the bounce point.  Then, the perturbations can be evolved to late times using \eqref{sc-hybrid} and its counterpart for tensor modes.

\subsection{Separate Universe Quantization}
\label{ss.sep}

The third approach to cosmological perturbation theory in LQC is based on the separate universe approximation used in cosmology to study long-wavelength perturbations \cite{Salopek:1990jq, Wands:2000dp}, where long-wavelength modes are those that satisfy $z_s''/z_s \gg k^2$.

Near the bounce, when LQC effects are important, $z_s''/z_s \sim a^2 \lp^{-2}$, and therefore the restriction of the separate universe approximation to long-wavelength modes corresponds to, near the bounce, only considering sub-Planckian modes.  So, the results in this section can safely be used, near the bounce, for sub-Planckian modes.  (Once LQC effects are negligible, the Mukhanov-Sasaki equation \eqref{class-ms} of general relativity can safely be used for all, short- and long-wavelength, sub-Planckian modes.)

The separate universe framework can be adapted to LQC to provide a full loop quantization of both the background and long-wavelength scalar perturbations.  The idea is simply to discretize a cosmological space-time with small perturbations into a lattice where each cell in the lattice is approximated to be homogeneous \cite{Bojowald:2006qu, Artymowski:2008sc, WilsonEwing:2011es, WilsonEwing:2012bx, Wilson-Ewing:2015sfx}.  In the separate universe framework, the discretization is chosen such that only long-wavelength modes are included, and in this case interactions between neighbouring cells are negligible and can safely be ignored.  Then, since each cell is homogeneous and uninteracting with other cells, a loop quantization is possible in each cell following the standard loop quantization techniques used on homogeneous space-times as reviewed in Sec.~\ref{ss.th}.

This quantization is particularly simple for scalar perturbations in the longitudinal gauge in which case the line element is
\be \label{long-g}
ds^2 = \bar N^2 (1 + 2 \psi) dt^2 + a^2(1- 2\psi) d \vec x^2,
\ee
where $\bar N$ is the background lapse and $\psi$ encodes the scalar perturbations, assuming vanishing anisotropic stress in the matter field.  The discretization of the space-time on a lattice gives cells that each have a local scale factor $a_i = a(1-\psi_i)$, and the long-wavelength perturbations are encoded in differences between the $a_i$ of different cells.

Clearly, in each homogeneous cell the line element is that of a spatially flat FLRW space-time (although with an unusual form for the lapse).  The loop quantization of the FLRW line element in each cell is straightforward and the result is a loop quantization of all of the degrees of freedom in the discretized version of \eqref{long-g}, namely both the background and the long-wavelength scalar perturbations \cite{Wilson-Ewing:2015sfx}.

Then, if the wave functions in each cell are sharply peaked, effective equations can be used in each cell to approximate the dynamics.  From these effective equations it is possible to extract the equations of motion for the background degrees of freedom, giving \eqref{fr1}--\eqref{cont} if $\bar N = 1$, and the dynamics of the perturbations are given by \cite{Wilson-Ewing:2015sfx}
\be \label{sep-v}
v_k'' - \f{z_s''}{z_s} \, v_k = 0.
\ee
Note that the form of $z_s$ in this equation is the result of a derivation starting from the full loop quantization in the separate universe approximation.  Thus, the separate universe approach suggests that $z_s = a \dot\phi / H$ is the correct term that should show up in the LQC Mukhanov-Sasaki effective equation, not $u = a \sqrt{3(1 + w_{eff})/8 \pi G}$ as used in the dressed metric approach (although how much of a quantitative difference this may make is not clear).

It is important to keep in mind that the effective equation \eqref{sep-v} is extracted from the quantum theory and only holds when the wave function in each cell is sharply peaked.  If quantum fluctuations are important, then this effective equation breaks down and it is necessary to instead use the full quantum equations of motion given in \cite{WilsonEwing:2012bx, Wilson-Ewing:2015sfx}.

So far, the loop quantization in the separate universe approximation has only been completed for scalar perturbations in the longitudinal gauge \cite{WilsonEwing:2012bx, Wilson-Ewing:2015sfx}.  In principle, it could be extended to tensor perturbations (as well as scalar perturbations in an arbitrary gauge) but then the line element would be significantly more complicated and the loop quantization of the space-time discretized on a lattice, while possible, will not be as simple as for \eqref{long-g}.

The separate universe approach to cosmological perturbation theory in LQC has the major advantage in that it is the only one that allows for a loop quantization of both background and perturbative degrees of freedom.  However, it has a number of drawbacks as well: it is only applicable to long-wavelength scalar perturbations, it requires a gauge-fixing of the scalar perturbations before quantization, and it can only be used if the matter fields have vanishing anisotropic stress.  In particular, it cannot be applied to cosmological perturbations with a wavelength comparable to (or smaller than) $\lp$ and so does not address the trans-Planckian problem.

Nonetheless, the equation of motion \eqref{sep-v} for scalar perturbations can be used in a number of interesting settings to calculate the evolution of long-wavelength cosmological perturbations through the LQC bounce.  This is particularly relevant for alternatives to inflation, like the matter bounce and ekpyrotic scenarios, where all of the observationally relevant modes today would have had a wavelength much larger than $\lp$ at the bounce point.

\section{Observational Effects}
\label{s.obs}

The predictions of LQC, just like general relativity, depend on the matter fields dominating the dynamics, and therefore LQC effects will vary depending on the cosmological scenario.  In addition, since the three approaches to cosmological perturbation theory in LQC outlined above have some differences, the predictions may also depend on the approach that is used.  If this is the case, then it will be necessary to determine which approach is more trustworthy.  However, the predictions seem to be mostly independent of the framework used, with one important exception, when the effective constraint approach is used outside of its domain of validity in inflationary models, that will be discussed below.

There is also the possibility that there may be some pre-bounce physics that could leave a signature in the CMB, for example circles in the CMB of low variance in the temperature as suggested in \cite{Gurzadyan:2010da}.  If this is indeed the case, the geometric characteristics of these circles can give important information into the pre-bounce era \cite{Nelson:2011gb}.  However, so far there is no sign of such circles \cite{Wehus:2010pj, Moss:2010py, Hajian:2010cy}, or of any other sign of structure from a pre-bounce epoch.

Therefore, the most promising direction for observational tests appears to be calculating how LQC modifies the predictions of various early universe models like inflation, the matter bounce scenario and the ekpyrotic universe, and check whether these effects could be detected in the CMB.

In cosmological perturbation theory, short- and long-wavelength modes evolve differently, since different terms in the Mukhanov-Sasaki equation dominate the dynamics: in the classical Mukhanov-Sasaki equation \eqref{class-ms}, for short-wavelength modes $k^2 \gg z_s''/z_s$ and for long-wavelength modes $z_s''/z_s \gg k^2$.  (The regime where both terms are comparable is typically very short for any given Fourier mode, and in fact can be approximated as a discontinuous transition with the requirement that $v_k$ and $v_k'$ be continuous at the transition time.)

The various approaches to cosmological perturbation theory in LQC predict a number of modifications to the classical Mukhanov-Sasaki equation, as reviewed in Sec.~\ref{s.perts}.  Depending on the cosmological dynamics, some modifications may leave traces in the CMB while others will not.

To be specific, holonomy corrections are most important near the bounce point when the radius of the space-time curvature is of the order of $\lp$, and therefore LQC corrections to the $k^2$ term will only affect modes that are trans-Planckian at the bounce point, while LQC corrections to $z_s''/z_s$ (that either modify the form of $z_s''/z_s$, or the background dynamics of $z_s$ via the LQC effective Friedmann equations) will only affect modes that are sub-Planckian at the bounce point.

While inverse triad effects are not as well understood as holonomy corrections in LQC, it is typically expected that they will become important for trans-Planckian modes (or perhaps within a few orders of magnitude of the Planck length) \cite{WilsonEwing:2011es}.  As a result, the LQC effects that are observationally relevant today depend on the post-bounce cosmological dynamics, and in particular whether modes that were trans-Planckian at the bounce are observable in the CMB today.

For example, in both the matter bounce and ekpyrotic scenarios the modes that are observed in the CMB today were all far away from the Planck scale, and therefore only LQC corrections to $z''/z$ in \eqref{class-ms} can leave any traces in the CMB.  On the other hand, for inflation with $\sim 70$ e-folds or more, at least some of the modes observed today were trans-Planckian at the bounce point and in this case LQC corrections to the $k^2$ term in \eqref{class-ms} may be found at small scales in the CMB.  In fact, for inflation with $\sim 80$ e-folds or more, all of the modes that were sub-Planckian at the bounce point are super-Hubble today, and in this case only LQC corrections to the $k^2$ term could potentially be observed today.

The results reviewed in the remainder of this section build on many earlier results, including the pioneering works \cite{Tsujikawa:2003vr, Zhang:2007bi, Bojowald:2007cd, Mielczarek:2008pf, Copeland:2008kz, Grain:2009kw} that first studied LQC effects on cosmological perturbations, but for reasons of space here I will focus only on the most recent results that represent the current understanding of the field.

\subsection{Inflationary Models}
\label{ss.inf}

In inflationary cosmological models, there is a long period of accelerated expansion of the universe, typically generated by a scalar field $\phi$ slowing rolling down its potential $V(\phi)$.  This inflationary phase generates from vacuum quantum fluctuations a nearly scale-invariant spectrum of primordial curvature perturbations, as observed in the CMB, and also of primordial tensor perturbations \cite{Baumann:2009ds}.  The most recent observations of the CMB strongly constrain the form of the inflationary potential \cite{Hinshaw:2012aka, Ade:2015xua}, and can also be used to test any LQC effects that could modify the standard inflationary predictions.

While LQC alone does not predict a long inflationary period, if there is an inflaton field (with a suitable potential $V(\phi)$) present then the probability is very high of there being a long era of slow-roll inflation \cite{Ashtekar:2009mm, Corichi:2010zp, Linsefors:2013cd}.  For typical solutions, the inflaton is kinetic-dominated before and during the bounce point due to the Hubble anti-friction term in the Klein-Gordon equation
\be
\ddot{\phi} + 3 H \dot{\phi} + \f{d V}{d\phi} = 0,
\ee
and this drives the inflaton far from the minimum of the potential.  Then, when expansion starts the Hubble term in the Klein-Gordon equation will act as a friction term on the inflaton, until the inflaton reaches the slow-roll regime.  Calculations show that for almost all solutions there will be at least $68$ e-folds of slow-roll inflation.

While most studies of inflation in LQC have focused on the simplest case where the inflationary potential is quadratic, $V(\phi) = m^2 \phi^2 / 2$ \cite{Ashtekar:2009mm, Agullo:2013ai}, other potentials have also been considered including the plateau potentials preferred by the latest observational data \cite{Bonga:2015xna}.  In any case, LQC effects are essentially independent of the specific form of the inflaton potential, for the reason that LQC effects become strong near the bounce and the inflaton field is typically kinetic-dominated during the bounce, as explained above, and so the dynamics are independent of the form of $V(\phi)$ when LQC effects are important.

On the other hand, LQC predictions do depend on the duration of inflation.  Recall that the dynamics of short- and long-wavelength perturbations (and LQC corrections thereof) are different.  So, the LQC effects that are potentially observable in the CMB depend on the length of inflation.  More specifically, LQC effects on long-wavelength perturbations may appear in the CMB only if there were at most $\sim 80$ e-folds of inflation (where the exact bound depends on the energy scale at which reheating occurs) \cite{Barrau:2013ula}, while LQC effects on short-wavelength perturbations may be visible in the CMB only if there were at least $\sim 70$ e-folds of inflation (since CMB temperature anisotropies have been measured over 8 e-folds of Fourier modes).

\subsubsection{Hybrid quantization}
\label{sss.hybrid}

In the hybrid quantization framework, there are only LQC corrections to the $\tilde u'' / \tilde u$ term via the LQC modifications of the background dynamics \eqref{fr1}--\eqref{cont}.  For this reason, at short scales the dynamics are the same as those of general relativity, and so in this case it may be possible to observe LQC effects only if there are less than $\sim 80$ e-folds of inflation.

In this approach quantum vacuum initial conditions are imposed at the bounce time, motivated by the heuristic picture of a repulsive force simultaneously generating the bounce and smoothing out the perturbations (as much as possible, given the quantum uncertainty relations).  However, there is an ambiguity in selecting a vacuum state for a quantum theory on a dynamical background.  So far, three possibilities have been considered in the hybrid quantization framework: (i) setting the vacuum state at the bounce point to be exactly the fourth-order adiabatic vacuum state at that time \cite{Agullo:2013ai}, (ii) requiring that oscillations in the perturbations be minimized at the initial time \cite{deBlas:2016puz}, and (iii) motivated by Penrose's hypothesis on the initial vanishing of the Weyl curvature \cite{Penrose:1979}, choosing the vacuum state so that the universe be as homogeneous and isotropic as possible at the bounce point, as permitted by the uncertainty relations \cite{Ashtekar:2016pqn}.  Note that since the vacuum state depends on the dynamics of the background space-time, and quantum gravity corrections are important during the bounce, these quantum gravity corrections will affect the choice of the initial vacuum state for the perturbations.

At a phenomenological level, differences between various choices regarding the initial vacuum state only become important at large scales since these differences vanish as $\tilde a / k \to 0$; in particular, at the bounce point the ambiguity in the choice of the vacuum state may have potentially observational consequences for sub-Planckian modes, but not for trans-Planckian modes.

For the choice (i) of the vacuum state, it is found that the power spectrum at large scales becomes oscillatory, with a frequency so rapid that the oscillations are not realistically observable.  What is observable is the average power spectrum, which is amplified compared to the standard inflationary prediction, and in addition, a positive running of the scalar spectral index is predicted at these scales, as well as a modification to the consistency relation of single field inflation at large scales \cite{Agullo:2013ai, Agullo:2015tca}.  The first prediction is not favoured by observations which find the power at large scales to be smaller than expected \cite{Bennett:2003bz, Ade:2013zuv}, rather than larger.  Of course, these three particular effects are observable only if the long-wavelength modes at the bounce point remain observable today, i.e., only if there were at most $\sim 80$ e-folds of inflation.

Interestingly, even if there were more than $\sim 80$ e-folds of inflation---in which case the amplified modes are today at super-horizon scales---then these modes could affect the observed power spectrum if there are strong correlations between the observable and the super-horizon modes, generated by the non-linearities in the dynamics of the cosmological perturbations.  For the vacuum choice (i), there is an amplification of power at large scales, as explained above, and non-Gaussianities (during the standard inflationary era) will induce correlations between the super-horizon modes and observable modes in the CMB.  Non-Gaussianities are strongest between the super-horizon modes and the CMB angular multipoles $\ell \lesssim 30$, and the effect of these non-Gaussianities on the CMB is a power asymmetry at large scales \cite{Agullo:2015aba}.  To be specific, this gives a dipole modulation in the power at large scales, in agreement with observations \cite{Ade:2015hxq}, with higher multipole modulation having a much smaller amplitude.  An important point is that the dipole modulation is strongest (and can explain observations only) if the amplified modes are only slightly super-horizon.

On the other hand, the choices (ii) and (iii) for the initial vacuum state both give a suppression of power at large scales \cite{Bonga:2015xna, Ashtekar:2016wpi}.  For choice (iii), this is a natural result of choosing the initial vacuum state so that perturbations (at large scales) are as small as allowed by quantum mechanics.  Interestingly, the suppression at large scales agrees quantitatively with observations \cite{Bennett:2003bz, Ade:2013zuv}.  This is a non-trivial result: while these initial conditions will clearly reduce power at large scales, this effect could a priori have been too large or too small to explain observations.  For choice (ii), this result is more surprising, but suggests that the choices (ii) and (iii) for the initial quantum vacuum may be related.

In addition, at least for the initial vacuum state (iii), the same effect modifies the E-mode polarization in a similar way and thus also suppresses the T-E and E-E power spectra at large scales \cite{Ashtekar:2016wpi} (with T denoting the temperature anisotropies and E the E-mode polarization).  These additional predictions will hopefully allow future observations of the CMB to differentiate between this scenario of primordial cosmology and other potential explanations for low power in the T-T power spectrum at large scales.  Finally, in this scenario the tensor power spectrum is also suppressed in the same fashion as the curvature power spectrum (although this effect will likely be harder to test for than suppression in the T-E and E-E power spectra).

These results raise three points.  First, the number of e-folds during inflation must be fine-tuned to approximately 72 e-folds for these LQC effects to occur at precisely the scales where the anomalies are observed in the CMB (i.e., $\ell \lesssim 30$ for the power deficit and $\ell < 64$ for the dipolar modulation in the power asymmetry).  Whether this fine-tuning can be justified as the result of pre-bounce dynamics or of some other feature that could limit the number of inflationary e-folds remains a challenge for future work.

Also, since the predicted power spectrum at large scales depends on the initial conditions imposed at the bounce point, it is important to address the ambiguity in the choice of the quantum vacuum state (at least in this relatively simple context of cosmological perturbation theory), perhaps in terms of the inputs used to define the `preferred' vacuum.  In particular, does LQC in some way suggest a new physical input that can be used to select a specific vacuum at a specific time?  The suggestions in \cite{deBlas:2016puz, Ashtekar:2016pqn} propose potential solutions to this problem.  On the other hand, if there is no canonical choice (based on some new physical principle) for the initial vacuum state, then the predictions of LQC in inflationary models at large scales will depend on the choice of the initial vacuum state (at least, if the initial conditions are to be imposed at the bounce point).  In this case, it may be possible to constrain the initial vacuum state by observations, but the theory will lose predictive power.

Finally, note that in the hybrid quantization approach it is assumed that the physical wavelength of cosmological perturbations can be arbitrarily small (and the results here clearly show that trans-Planckian modes are not problematic in this setting).  For more on this assumption, see Sec.~\ref{ss.planck}.

\subsubsection{Effective constraint approach: Holonomy corrections}
\label{sss.hol}

LQC effects in an inflationary background have also been studied in the effective constraint approach to cosmological perturbation theory, taking into account the effect of holonomy corrections to tensor \cite{Linsefors:2012et} and scalar \cite{Bolliet:2015bka} modes.  In this case there are LQC corrections to the dynamics of both short- and long-wavelength cosmological perturbations.

At large scales, there is a slight `bump' of increased power near the Fourier mode $k_o$ whose wavelength was $\sim \lp$ at the bounce time, and then the power is slightly suppressed at scales larger than $k_o$ \cite{Barrau:2013ula}.  Further work is needed to determine whether this LQC effect could account for the observed low power at large scales \cite{Bennett:2003bz, Ade:2013zuv}, although it would certainly require fine-tuning to obtain the correct number of e-folds for the suppression to appear at the correct scale.  The same bump and suppression of power at large scales is also predicted for the tensor modes (although of course with a smaller amplitude).

At short scales (i.e., the Fourier modes that are trans-Planckian during the bounce) there is an exponential growth in the amplitude of the tensor and scalar perturbations during bounce \cite{Linsefors:2012et, Bolliet:2015bka}.  Clearly, this amplified power at small scales is not observed and this rules out this type of LQC effect \cite{Bolliet:2015raa}.  However, note that this exponential growth only occurs for trans-Planckian modes, which is precisely where the effective formalism breaks down since quantum fluctuations are no longer negligible at the Planck scale (and quantum fluctuations being negligible is a key assumption in this version of the effective constraint framework; in principle, quantum fluctuations could be included in an appropriate extension of the effective constraints).  Therefore, this prediction is a result of this approach being applied outside of its domain of validity and one should not be surprised that the effective constraint approach breaks down in this regime.  Note that the predictions for long-wavelength perturbations, which are obtained within the regime of validity of the theory, are not ruled out by observations.  Therefore, the lesson here appears to be not that the effective constraint approach is wrong, but rather that it is important that it be used only in its regime of validity, i.e., for sub-Planckian modes only.

Finally, as is clear from the discussion here and in Sec.~\ref{sss.hol}, this effective constraint approach to cosmological perturbation theory in LQC and the hybrid quantization framework give different predictions in the inflationary setting for short-wavelength modes.  This is due to this effective constraint approach being used outside of its domain of validity (i.e., for trans-Planckian modes).  The hybrid quantization approach, on the other hand, does not break down for trans-Planckian modes, and therefore the results obtained for trans-Planckian modes using the hybrid quantization are more reliable.

\subsubsection{Effective constraint approach: Inverse triad effects}
\label{sss.inv}

The effect of inverse triad corrections on tensor \cite{Grain:2009cj, Bojowald:2010me} and scalar \cite{Bojowald:2010me, Bojowald:2011iq} perturbations has also been studied in the effective constraint approach.  For the LQC of homogeneous cosmological space-times, inverse triad effects only become important when length scales approach the Planck scale (or come within a few orders of magnitude if inverse triad effects are strong).  This usually does not happen in homogeneous LQC since the bounce occurs when the space-time curvature is Planck-scale, and in typical solutions the physical volume of the space-time at the bounce point will be much larger than $\lp^3$.  However, this is no longer the case for cosmological perturbations: in inflation with more than $\sim 70$ e-folds, some Fourier modes of the perturbations (that are observationally relevant today) will have a wavelength comparable to the Planck scale at early, near-bounce times.  Therefore, cosmological perturbation theory is a particularly promising arena to study inverse triad effects and constrain them via observations.

For both tensor and scalar modes, the inverse triad effects studied so far generate a larger-than-expected running of the spectral index by enhancing power at large scales \cite{Grain:2009cj, Bojowald:2011iq}.  (The surprising result that these inverse triad effects modify the spectrum at large scales rather than at short scales is not yet fully understood.)  In inflationary models with less than $\sim 80$ e-folds, this enhancement would be present at scales observable today, and in this case the latest observational bounds on the running of the spectral index to be at most of the order $10^{-2}$ \cite{Ade:2015xua}, as well as the observed smaller-than-expected---rather than larger---amplitude of the scalar power spectrum at large scales \cite{Bennett:2003bz, Ade:2013zuv} strongly constrain the amplitude of inverse triad effects in cosmological perturbation theory.

\subsection{Matter Bounce Scenario}
\label{ss.mb}

An alternative to inflation is the matter bounce scenario, where vacuum fluctuations in curvature and tensor perturbations become scale-invariant in a contracting FLRW space-time where the matter content has vanishing pressure.  Then, if a bounce can be generated to provide a non-singular transition from contraction to expansion, these scale-invariant perturbations provide suitable initial conditions for the expanding universe that can explain the (near) scale-invariance observed in the CMB, under the assumption that the bounce does not modify the power spectrum \cite{Brandenberger:2012zb, Cai:2016hea}.

It is immediately clear that the quantum gravity effects of LQC can generate the bounce that is required for this scenario to be viable, and furthermore it is possible to explicitly calculate the evolution of the curvature perturbations across the bounce to verify that they do in fact remain (nearly) scale-invariant.  This calculation can be done using the separate universe approach, since all of the observationally relevant modes are in the long-wavelength limit throughout the bounce.  (The analogous calculation for tensor modes can be done using the results of the effective constraint approach.)

There are two main results: first, both the curvature and tensor modes remain scale-invariant throughout and after the bounce, and second, the amplitude of the tensor modes are typically suppressed by LQC effects during the bounce, in some cases significantly \cite{WilsonEwing:2012pu}.  The suppression depends on the equation of state of the matter field during the bounce; the closer the equation of state is to zero, the more the tensor-to-scalar ratio will be suppressed during the bounce \cite{Wilson-Ewing:2015sfx}.  In particular, if the dominant matter field during the bounce is radiation (as was the case in the early universe), then the equation of state is $\omega = 1/3$ and the tensor-to-scalar ratio will be suppressed by a factor of 1/4 during the bounce \cite{Cai:2014jla}.

Therefore, not only does LQC provide the bounce required by the matter bounce scenario and preserves the scale-invariance of the perturbations across the bounce, but it may also leave a quantum gravity signal in the CMB, namely a smaller-than-expected tensor-to-scalar ratio.  This effect could allow observations to distinguish between a matter bounce scenario with an LQC bounce or with a bounce generated by other physics, e.g., by a matter field violating energy conditions.  Furthermore, the amplitude of the suppression of the tensor-to-scalar ratio depends on the dominant matter field during the bounce, and so if this effect is indeed measured in the CMB in the future, giving evidence of an LQC bounce, it would also provide important information concerning the dominant matter field during the bounce.

\subsection{Ekpyrotic Universe}
\label{ss.ekp}

Ekpyrotic scenarios have also been considered in the context of LQC.  The ekpyrotic universe is cyclic, with a scalar field whose potential allows the scalar field to act as dark energy in an expanding universe, cause a recollapse after a long dark-energy-dominated era, and act as an ultra-stiff fluid during contraction.  Due to the ultra-stiff behaviour of the scalar field during contraction, not only do anisotropies remain small at all times, but also vacuum entropy perturbations become scale-invariant.  These scale-invariant entropy perturbations can then act as a source to generate scale-invariant curvature perturbations, and if the contracting era is followed by a bounce, these scale-invariant perturbations provide good initial conditions to generate the CMB, if they are not modified during the bounce \cite{Lehners:2008vx}.

One of the main challenges of the ekpyrotic scenario is to generate the bounce to pass from the contracting to the expanding phase.  As LQC automatically replaces the big-crunch singularity by a non-singular bounce, it is natural to consider a realization of the ekpyrotic universe with an LQC bounce.  This possibility has been explored, both at the homogeneous level \cite{Cailleteau:2009fv}, and also studying the dynamics of perturbations as they cross the bounce \cite{Wilson-Ewing:2013bla}.  Since the observationally relevant modes today would all have been far from the Planck scale during the bounce, the separate universe approach to cosmological perturbation theory in LQC can safely be used here.

Using \eqref{sep-v}, it is easy to check that if the curvature perturbations are scale-invariant before the bounce (as sourced by the scale-invariant entropy perturbations), then these curvature perturbations remain scale-invariant during and after the bounce \cite{Wilson-Ewing:2013bla}, giving a viable realization of the ekpyrotic scenario.  On the other hand, in the absence of entropy perturbations, the curvature perturbations after the bounce have a blue spectrum and so this possibility is ruled out.  Thus, entropy perturbations play an essential role in ekpyrotic models.  Also, while LQC can naturally provide the bounce the ekpyrotic scenario needs, it does not seem to leave any imprint on the cosmological perturbations.  In other words, in the ekpyrotic scenario there does not appear to be a way to differentiate between a bounce caused by LQC, or a bounce caused by, e.g., a violation of the weak energy condition.

\section{Cosmological Constant}
\label{s.lambda}

LQC is compatible with a cosmological constant $\Lambda$, whether it is positive \cite{Kaminski:2009pc, Pawlowski:2011zf} or negative \cite{Bentivegna:2008bg}.  The only constraint is that $\Lambda$ must be smaller than $8 \pi G \rho_c$ for the quantum theory to be non-trivial and have a good semi-classical limit; this is necessary so that the energy density associated to the cosmological constant is smaller than $\rho_c$, the critical energy density of LQC.

Therefore, LQC can easily include a small positive cosmological constant, as observations appear to require.  However, LQC does not address what is often called the `cosmological constant problem' which asks for an explanation from fundamental physics of why $\Lambda$ is so small compared to the Planck scale.

Rather, in LQC (and, more generally, in LQG) the point of view is typically that the cosmological constant is a constant of nature like Newton's gravitational constant or Planck's constant.  If this point of view is correct, then $\Lambda$ should be measured through some experiments and/or observations, and this measurement will determine the value of $\Lambda$; it does not arise as the result of a more fundamental calculation \cite{Bianchi:2010uw, Bianchi-Rovelli-Kolb}.  (While it has been suggested that the value of $\Lambda$ may run \cite{Niedermaier:2006wt}, this possibility has not yet been investigated in LQC.)

\section{Open Questions}
\label{s.open}

Loop quantum cosmology is by now a mature field where predictions can be calculated explicitly for a variety of cosmological scenarios, as reviewed above.  However, there still remain some important open problems that need to be addressed.  Here I briefly discuss what are in my opinion three of the main open problems in LQC: determining its relation to LQG, the trans-Planckian problem, and a possible signature change from a Lorentzian to a Euclidean space-time around the bounce point.

\subsection{Relation to Loop Quantum Gravity}
\label{ss.lqg}

While LQC uses the same variables and quantization techniques as loop quantum gravity, it has not been derived from LQG.  The reason for this is that in LQC the symmetries of the cosmological space-times of interest are imposed before quantization, rather than the reverse.  Since quantization and symmetry reduction do not necessarily commute, it is important to understand the relation between LQC and the cosmological sector of LQG.  In particular, it is necessary to determine if there are any important LQG effects that are not captured by LQC.

First, at the kinematical level, it has been shown that the LQC kinematical Hilbert space can be embedded in the LQG Hilbert space \cite{Engle:2007zz, Brunnemann:2007du, Engle:2013qq} and that furthermore the requirement of invariance under the residual diffeomorphisms allowed by the gauge-fixing \eqref{vars} uniquely selects the representation of the (symmetry-reduced) holonomy-flux algebra used in LQC \cite{Ashtekar:2012cm, Engle:2016hei, Engle:2016zac}.  Based on these results the relation between the two theories at the kinematical level is quite well understood.

Less is known at the dynamical level.  A number of approaches have been developed to address this problem, including spin foam cosmology \cite{Bianchi:2010zs, Bianchi:2011ym, Rennert:2013pfa, Rennert:2013qsa, Vilensky:2016tnw} and quantum-reduced loop gravity, whether based on one node representing all of space \cite{Bodendorfer:2014vea, Bodendorfer:2015hwl} or many nodes in a lattice \cite{Lin:2011vz, Alesci:2012md, Alesci:2013xd, Alesci:2016rmn, Bodendorfer:2016tky}.  The quantum Friedmann dynamics can be extracted in these approaches and the correct classical limit is recovered if an important lesson from LQC is used: the length of the holonomies constituting the field strength operator in the Hamiltonian constraint must depend on the densitized triad operator, as discussed in Sec.~\ref{ss.th} below Eq.~\eqref{field}.  While these results are very encouraging, they rely in an essential way on input from LQC itself, and it would be nice to go beyond them.

One potential way forward in this direction is offered by the suggestion that condensate states in group field theory (a second-quantized reformulation of LQG) may correspond to the cosmological sector of LQG \cite{Gielen:2013naa}.  Interestingly, for a group field theory corresponding to gravity coupled to a massless scalar field, it is possible to extract the cosmological dynamics of a certain type of condensate states through an appropriate coarse-graining, with the result that the big-bang singularity is generically resoved and replaced by a bounce \cite{Oriti:2016qtz}.  Furthermore, for a particularly simple family of condensate states, the cosmological dynamics are almost exactly the effective Friedmann equations of LQC \eqref{fr1}--\eqref{cont}.  Note that these results are obtained without requiring any direct input from LQC.  Finally, for some group field theory models, a low spin regime emerges at low curvatures \cite{Gielen:2016uft}---precisely in accordance with heuristic expectations coming from the theoretical underpinnings of LQC \cite{Ashtekar:2006wn, Ashtekar:2009vc, Pawlowski:2014nfa}---and the dynamics of the condensate states can also naturally generate accelerated expansion and/or the recollapse at large scales necessary for a cyclic universe \cite{deCesare:2016rsf}.  For a more detailed review of the group field theory approach to cosmology, see \cite{Gielen:2016dss}.

\subsection{Trans-Planckian Problem}
\label{ss.planck}

A requirement for any theory of quantum gravity is that it must predict whether (cosmological) perturbations can have a wavelength shorter than the Planck length or not, and if so the theory must provide the equations of motion for these trans-Planckian modes, which may include important quantum gravity corrections.

So far, as already mentioned in Sec.~\ref{s.perts}, this problem has not been fully addressed in LQC: trans-Planckian modes are outside the regime of validity of both the effective constraint and separate universe approaches to cosmological perturbation theory in LQC.  And while trans-Planckian modes are safely included in the hybrid quantization approach to cosmological perturbation theory in LQC, this is due to the assumption that there is no minimal length in LQG rather than being the result of a calculation.

At present, it is not known whether there is a minimal non-zero eigenvalue to the length operator in LQG, due to the complexity of the length operators proposed in the LQG literature \cite{Thiemann:1996at, Bianchi:2008es, Ma:2010fy}.  To clarify the situation concerning trans-Planckian perturbations in LQC, it is important to determine the spectrum of the length operator.  If it is found that there exist arbitrarily small eigenvalues of the length operator, this will support the hypothesis used in the hybrid quantization.  Otherwise, it may be necessary to correct the results obtained so far by introducing a minimal length cut-off in an appropriate manner.

\subsection{Signature Change?}
\label{ss.sig}

An intriguing result in the effective constraint approach to cosmological perturbation theory in LQC when including holonomy corrections is that the constraint algebra changes.  To be specific, while the Poisson brackets of the diffeomorphism constraint with itself and the Poisson bracket between the diffeomorphism and scalar constraints remain the same, the Poisson bracket of the scalar constraint with itself becomes \cite{Cailleteau:2011kr}
\be \label{mod-const}
\{\mH[N], \mH[\tilde N] \} = \left( 1 - \f{2 \rho}{\rho_{c}} \right) \mH_a[(N \partial^a \tilde N - \tilde N \partial^a N)],
\ee
where $\mH[N] := \int N \mH$.  This is the usual relation in general relativity, modified by the prefactor of $(1 - 2 \rho / \rho_c)$.  This modification signals a strong departure from general relativity in the Planck regime: the standard constraint algebra of general relativity uniquely determines the Hamiltonian and diffeomorphism constraints (up to the values of $G$ and $\Lambda$), under the assumptions that general relativity is second-order in derivatives and that $\mH$ and $\mH_a$ are respectively the generators of time-like and space-like diffeomorphisms \cite{Hojman:1976}.  Furthermore, even higher-derivative theories of gravity also have the same constraint algebra \cite{Deruelle:2009zk}.  So, the corrections found in the effective constraint approach cannot come from higher space-time curvature terms.

Furthermore, in the modified constraint algebra \eqref{mod-const} the prefactor becomes negative near the bounce, in the regime that $\rho_c / 2 < \rho \le \rho_c$, and at the bounce point where $\rho = \rho_c$ the prefactor is $-1$.  It has been suggested that this may correspond to a signature change from a Lorentzian space-time to a Euclidean space-time \cite{Bojowald:2011aa, Mielczarek:2012pf, Bojowald:2012ux, Bojowald:2015gra}, for the reason that the only difference between the constraint algebras for Lorentzian and Euclidean geometries is an overall sign in the Poisson bracket of the scalar constraint with itself, and here the prefactor goes from 1 in the classical limit to $-1$ at the bounce point.  In addition to this, the equations of motion \eqref{ef-sc}--\eqref{ef-te} for the scalar and tensor perturbations obtained in the effective constraint framework become elliptic around the bounce point, also suggesting a signature change.

This is an interesting proposal, and if true would suggest an unexpected convergence between LQC and other a priori completely unrelated approaches to quantum cosmology based on quantum tunneling from a Euclidean instanton \cite{Vilenkin:1982de, Hartle:1983ai}.  Furthermore, a signature change in the Planck regime would suggest a new way to impose initial conditions for cosmological perturbations and could generate new observational effects \cite{Mielczarek:2012tn, Gomar:2014wta, Mielczarek:2014kea, Schander:2015eja, Barrau:2016sqp}.  As an aside, note that (contrary to what has occasionally been stated in the literature) even if there is a signature change around the bounce point, it is possible to speak of a bounce in a relational sense.  Since the matter fields near the bounce are kinetic dominated, they will behave monotonically and so will be good relational variables during the bounce.  Therefore, even if there is no time coordinate in the vicinity of the bounce due to the Euclidean nature of space-time there, there nonetheless exists a well-defined relational framework wherein dynamics (with respect to the relational scalar field) are well-defined, including around the bounce point.

However, not all evidence points in the direction of a signature change at the bounce.  First, the equations of motion for perturbations become elliptic near the bounce also in other bouncing cosmologies (see, e.g., \cite{Cai:2012va}) where the space-time is clearly Lorentzian at all times.  Therefore, the fact that the equations of motion for the perturbations become elliptic, on its own, is not enough to show that the space-time becomes Euclidean.  (Note also that the exponential instability in the equations of motion only affects trans-Planckian modes \cite{Bolliet:2015bka}, precisely where these equations of motion break down.)  Second, the hybrid quantization approach to cosmological perturbation theory sees no evidence of a signature change.  (The separate universe approach cannot address this question since its key approximation is to ignore interactions and hence spatial derivatives, and in that limit $\{\mH[N], \mH[\tilde N]\} = 0$.)

Furthermore, if the signature change proposal is correct, then since the Poisson bracket \eqref{mod-const} is the same as for Euclidean general relativity at the bounce point (as are the other two Poisson brackets in the constraint algebra), and since the effective constraints are second-order in derivatives, it should follow immediately from the results of \cite{Hojman:1976} that the LQC Hamiltonian constraint at the bounce point should also be that of Euclidean general relativity.  However, this is not the case, as can easily be checked.  It is not immediately clear why the results of \cite{Hojman:1976} do not hold in this case, although there are several possibilities: (i) the results of \cite{Hojman:1976} are derived using the spatial metric and its conjugate momentum rather than the Ashtekar-Barbero variables, perhaps the results do not hold for a different choice of elementary variables; (ii) perhaps in the effective constraint approach the constraints can no longer be interpreted as generators of diffeomorphisms (although this would be problematic since there would no longer exist a clear space-time interpretation); (iii) another possibility is that the phase space---assumed to be unchanged in the effective constraint approach---may in fact need to be enlarged if quantum corrections add higher derivative terms to the action; (iv) finally, perhaps for the results of \cite{Hojman:1976} to hold, the numerical prefactor to \eqref{mod-const} must be exactly $1$ (or $-1$ for Euclidean space-times), and that even infinitesimal departures from this are not allowed---but if this last possibility is indeed the case, then clearly the modified term \eqref{mod-const} in the constraint algebra does not suggest that space-time becomes Euclidean, even in a neighbourhood of the bounce point, since in any neighbourhood of the bounce point the prefactor is not everywhere exactly $-1$.  Obviously, an important open problem is to understand precisely why the results of \cite{Hojman:1976} are not applicable to LQC, in the sense described in this paragraph.  Until this last point is understood, it will not be clear whether there truly is a signature change in LQC.

Finally, the possibility of a signature change occurring around the LQC bounce point has been thrown into further doubt by some recent results of the effective constraint approach to cosmological perturbation theory based on self-dual LQC.  While the version of homogeneous LQC based on self-dual variables is qualitatively similar to standard LQC insofar as it also predicts that the big-bang singularity is resolved by a cosmic bounce \cite{Achour:2014rja, Wilson-Ewing:2015lia, Wilson-Ewing:2015xaq}, important qualitative differences arise when perturbative degrees of freedom are included: in the effective constraint approach based on self-dual variables, the constraint algebra is unchanged from that of general relativity \cite{BenAchour:2016leo}.  As a result, for self-dual variables it may be possible to interpret the modifications to the effective constraints as higher space-time curvature corrections, something that was impossible for Ashtekar-Barbero variables.  In any case, there is no indication of signature change in self-dual LQC.

To summarize, although there are intriguing results in the effective constraint approach to cosmological perturbation theory that may appear to hint at a signature change, further work is needed to confirm or rule out this possibility.

\section{Summary}
\label{s.disc}

The main prediction of loop quantum cosmology is that the big-bang singularity is resolved due to quantum gravity effects and is replaced by a cosmic bounce.  In addition, frameworks have been developed to study cosmological perturbation theory in LQC.  The quantitative predictions of LQC (just like classical general relativity) depend on the matter fields dominating the dynamics and therefore the predictions of LQC, including those concerning the CMB, depend on the cosmological scenario.  In particular, effects in inflation and the matter bounce scenario are different.

In inflation with $\sim 72$ e-folds starting from the bounce point, LQC can generate a power spectrum of the curvature perturbations with less power at large scales.  (Less e-folds, for this choice of the initial vacuum state for the perturbations, is ruled out observationally, while if there were more e-folds this effect would be confined to super-horizon scales and would not be observable.)  In this case, the same power suppression effect is predicted to occur also in the T-E and E-E correlation functions as well as in the power spectrum of tensor modes.

In the matter bounce scenario, LQC can suppress the tensor-to-scalar ratio during the bounce.  The precise numerical factor by which the tensor-to-scalar ratio is suppressed depends on the dominant matter field during the bounce; for example, the tensor-to-scalar ratio is suppressed by a factor of $1/4$ if the bounce is radiation-dominated.  Measuring this suppression factor would therefore provide important information about the physics of the bounce.

In short, while there do remain some important open questions, loop quantum cosmology is now a mature field where it is possible to explicitly calculate predictions in a number of interesting cosmological settings that can realistically be tested by observations of the CMB.

\bigskip

\acknowledgments

I would like to thank Ivan Agull\'o, Aur\'elien Barrau, Martin Bojowald, Guillermo Mena Marug\'an and Parampreet Singh for helpful discussions.


\raggedright

\end{document}